\documentclass[a4paper,12pt,twoside]{article}

\usepackage{latexsym}
\usepackage{mathtools}   
\usepackage{amssymb}           
\usepackage{amstext}               
\usepackage{amsfonts}            
\usepackage{icomma}            
\usepackage{enumitem}      
\usepackage{array}         
\usepackage{multirow}
\usepackage{graphicx,subfigure}
\usepackage{fancyhdr}
\usepackage[table]{xcolor}

\usepackage[T1]{fontenc}
\usepackage[utf8]{inputenc}
\usepackage[top=0.7in, bottom=0.75in, left=0.625in, right=0.625in]{geometry} 
\usepackage[backend=biber, style=nature, citestyle=numeric-comp, sorting=none]{biblatex}
\usepackage{float}
 \DeclareUnicodeCharacter{0301}{\'{ }}

\fancypagestyle{fancyarticle}{
    \fancyhf{}
    \fancyfoot[CO, CE]{\thepage}
        \fancyhead[LE]{\footnotesize Density distribution of photospheric vertical electric currents in flare active regions of the Sun}
        \fancyhead[RE]{\footnotesize Zimovets et al.}
                \fancyhead[LO]{\footnotesize Density distribution of photospheric vertical electric currents in flare active regions of the Sun}
        \fancyhead[RO]{\footnotesize Zimovets et al.}
    }
    
\title{Density distribution of photospheric vertical electric currents in flare active regions of the Sun}
\author{Zimovets et al.}

\begin{document}
\pagestyle{fancyarticle}
\thispagestyle{plain}
\fontfamily{ptm}\selectfont
\noindent
\LARGE{Density distribution of photospheric vertical electric currents in flare active regions of the Sun}\\
\line(1,0){505}\\
\vspace{0.1cm}\\
\normalsize
\textit{I.V. Zimovets$^1$ , A.B. Nechaeva$^{1,2}$ , I.N. Sharykin$^1$ , W.Q. Gan$^3$}\\
\\
$^1$ Space Research Institute of the Russian Academy of Sciences, 117997, Profsoyuznaya Str. 84/32,
Moscow, Russia; e-mail: \textit{ ivanzim@iki.rssi.ru ; ivan.sharykin@phystech.edu}\\
$^2$ Moscow Institute of Physics and Technology, 141701, Institutskiy per., 9,
Dolgoprudny, Moscow Region, Russia; e-mail: \textit{nechaeva.ab@phystech.edu}\\
$^3$ Purple Mountain Observatory of the Chinese Academy of Sciences, 210034, No.8 Yuanhua Road,
Qixia District, Nanjing, China; e-mail: \textit{wqgan@pmo.ac.cn}\\
\\
\textbf{\large{Annotation}}\\
Solar active regions contain electric currents. Information on the distribution of currents is important for understanding the processes of energy release on the surface of the Sun and in the overlying layers. The paper presents an analysis of the probability density function (PDF) of the absolute value of the photospheric vertical electric current density ($|j_z|$) in 48 active regions before and after flares in 2010--2017. Calculation of $|j_z|$ is performed by applying the differential form of Ampere’s circuital law to photospheric vector magnetograms obtained from
observations of the Helioseismic and Magnetic Imager (HMI) instrument onboard the Solar Dynamics Observatory (SDO). It has been established that for the studied active regions PDF($|j_z|$) is described by the Gauss function in the low-$|j_z|$ region ($|j_z |<10110 \pm 1321$
statampere/cm$^2$) and the decaying power-law function in the region of higher $|j_z|$ values. Also, for some active regions PDF($|j_z|$) can be described by the special kappa-function. The distributions
of the parameters of the approximating functions are obtained using the least squares method. The average absolute value of the power-law function index is $3.69 \pm 0.51$, and $3.99 \pm 0.51$ of the
kappa-function. No systematic changes in parameters during the flares are detected. An explicit connection between the parameters and the flare X-ray class, as well as with the Hale magnetic class of the active regions, is not found. Arguments are presented in favor of the suggestion that the Gaussian distribution in the low-value region of PDF($|j_z|$) represents noise in the data, while the power-law “tail” reflects the nature of electric currents in the solar active regions.

\section{Introduction}
Solar magnetic fields determine solar activity, coronal heating, and acceleration of the solar wind. Magnetic fields can be measured routinely on the photosphere. Based on the measurements and theories, it was established that active regions are penetrated by fields concentrated in magnetic flux tubes \cite{1994ASSL..189.....S}, \cite{2015ASSL..417.....R}. On the basis of Ampère's circuital law, it was found that electric currents can flow along these tubes \cite{severny}, \cite{2018GMS...235...43F}. As vector magnetograms are only available for a thin layer, one usually can obtain only the vertical component of electric current on the photosphere ($j_z$). However, several attempts to estimate the horizontal component have also been made \cite{severny}, \cite{2010ApJ...721L..58P}, \cite{2017Ap.....60..544F}.\\

It is essential to study electric currents in active regions because of some reasons described in \cite{2018GMS...235...43F}, \cite{2018GMS...235..391S}, \cite{zaitsevstepanov}. Firstly, free magnetic energy required for the solar activity such as coronal jets, flares and coronal mass ejections (CMEs), is contained in electric currents. Dissipation of electric currents, both longitudinal to the magnetic field and as current sheets, leads to the transformation of free energy into kinetic energy of plasma and populations of accelerated particles, energy of electromagnetic radiation in a wide range of the spectrum, and into the energy of various waves. Secondly, Joule dissipation can affect energy balance in different solar atmosphere layers. Thirdly, presence of currents can affect the nature of propagation and dissipation of Alfvén waves in active regions, which may be important for the problem of coronal heating and solar wind acceleration.\\

In general, it is known that there is a connection between $j_z$ and flare productivity of active regions \cite{severny}, \cite{1990IzKry..81....8A,2017SoPh..292..159K, 2018Ge&Ae..58.1129F}. Detailed studies are needed to find out exactly how $j_z$ is associated with flares. Traditional approach in studying electric currents in active regions is to construct density maps of $j_z$ on the photoshere based on vector magnetograms and to analyze relationship between  spatial structure of $j_z$ and electromagnetic radiation sources of solar activity processes. In particular, a number of studies have been carried out on the relationship of radiation sources (microwave, H-alpha, ultraviolet, x-ray) of flares with photospheric $j_z$, and no unambiguous spatial relationship was found \cite{1968SoPh....3..282M,1990SvA....34..656R,1990IzKry..81....3A,1991RSPTA.336..381C,1997ApJ...482..490L,2015A&A...580A.106M,2014ApJ...788L..18S,2015ApJ...807..102S,2016ARep...60..939L,2017AGUFMSH41A2751Z}.It is found that the flare sources tend to appear on the edges of regions of strong $j_z$ and to avoid their local maxima. Since the connection between photospheric $j_z$ and flare sources is ambiguous, it is worth trying to find some patterns with the use of statistical analysis.\\

However, despite a fairly large number of articles on the study of $j_z$, we are not aware of the works on systematical study of  probability density function (PDF) of the density $j_z$ on the photosphere and relationship of its features with energy release processes in active regions. For example, it was done for the density of electric currents in the corona based on the modeling and extrapolation of magnetic field from the photosphere for single active regions \cite{2016PASJ...68..101K}, \cite{2016A&A...596A..56M}. PDF of density of electric currents was found to be able to be represented by a power function or a double power function (with a break). It is worth mentioning, however, that the results of extrapolation of magnetic fields are controversial. They depend on the method and quality of boundary data. In the paper \cite{2016PASJ...68..101K} an example of PDF($|j_z|$) for one active region (AR12158; SOL2014-09-10) was also given, which was visually different from PDF for coronal currents, but quantitative analysis of PDF($|j_z|$) was not done, and its form was not studied. \\

The aim of this work is to study the form of PDF($j_z$) for a set of active regions producing flares. In addition, we will check whether there are systematical differences in PDF($j_z$) before and after flares, and whether there is a correlation between parameters of PDF($j_z$) and both X-ray flare class and Hale magnetic class of active regions \cite{1919ApJ....49..153H}.

\section{Data and methods}

First of all, we need to mention that the idea of this work arose during the statistical study of relationship between hard X-ray flare sources and photospheric vertical electric currents $j_z$ \cite{2017AGUFMSH41A2751Z} when we estimated the error of $j_z$ by calculating PDF($j_z$) based on the data obtained from the Helioseismic and Magnetic Imager (HMI) \cite{2012SoPh..275..207S} on board the Solar Dynamics Observatory (SDO). It explains our choice of active regions for this study. 48 active regions, where the flares happened near the center of the solar disk (helioprojective
coordinates of the flare X-ray sources are $-600'' < (x_f,y_f) < 600''$, i.e the heliographic longitude and latitude ranges from $-40^{\circ}$ to $40^{\circ}$) and where it was possible to determine coordinates of the hard X-ray sources in the energy range of 50--100 keV based on the data from the Reuven Ramaty High-Energy Solar Spectroscopic Imager (RHESSI) \cite{2002SoPh..210....3L}, were selected for the time period from 2010 May to 2017 October. This time period was chosen because of the simultaneous observation of the Sun by SDO and RHESSI. Information about the studied active regions and flares will be published in the upcoming paper.\\

In this study we use vector photospheric magnetograms obtained from HMI/SDO provided for public (http://jsoc.stanford.edu/) as the Spaceweather HMI Active Region Patch (SHARP) product \cite{2014SoPh..289.3549B}, \cite{2014SoPh..289.3483H}. Standard data files of the “hmi.sharp\_cea\_720s.fits” type with the 12 minutes time step are used. With the use of the special algorithm, for each time interval a finite region (patch) was cut from the HMI field-of-view. Each patch corresponds to an active region and its vicinity and is determined by its HARPNUM number. In these files magnetic field vector $\Vec{B}$ ($B_r, B_\phi, B_\theta$) in the spherical coordinate system is projected onto Lambert cylindrical grid ($x = \phi, y = (180^{\circ}/\pi)\sin\theta$) with the equal cell square of $1.33 \cdot 10^5$ km$^2$ on the photosphere (Lambert Cylindrical Equal-Area projection, CEA) \cite{2002A&A...395.1077C}, \cite{2006A&A...449..791T}. $180^\circ$-uncertainty of the transverse to the line-of-sight component of magnetic field ($B_\perp$) is eliminated in that data. Using the “WCS” software package as part of “SolarSoft”, CEA coordinates were transformed into the Stonyhurst heliographic coordinate system and then into the spherical coordinate system centered on the center of the Sun.\\

Density of photospheric vertical electric currents are calculated in the spherical coordinate system with the use of Ampère's circuital law in differential form:
\begin{equation}
j_{z}=j_{r}=\frac{c}{4 \pi \mu}(\nabla \times \Vec{B})_{r} \approx \frac{c}{4 \pi} \frac{1}{R_{s} \sin \theta}\left(\frac{\Delta B_{\varphi}}{\Delta \theta} \sin \theta+B_{\varphi} \cos \theta-\frac{\Delta B_{\theta}}{\Delta \varphi}\right),
\end{equation}

where $c$ is the speed of light in vacuum, $R_s$ is radius of the Sun, and the magnetic permeability  $\mu = 1$. For each active region we constructed two maps of $j_z$: right before the onset of the flare soft X-ray rise and after the end of the flare impulsive phase, when hard X-ray emission ($> 25$ keV) drops to the pre-flare level. This gives us the opportunity to examine possible changes in PDF($j_z$) because of flare, eliminating variations of $j_z$ that may occur as the result of disturbances of the photosphere made by beams of accelerated particles, hydrodynamic flows, shock waves, and flows of electromagnetic radiation in the impulsive phase of a flare \cite{2019arXiv190610788S}. Examples of obtained maps of photospheric $j_z$ for two active regions, NOAA 12172 SOL2014-09-23 and 11263 SOL2011-08-03, before the flares are shown in Figure \ref{fig1:a} and \ref{fig1:b} respectively. For visualization $j_z$ maps were transferred to a homogeneous grid constructed in the helioprojection coordinate system.\\

Then histograms of $j_z$ were made with the bin size of 2500 statampere/cm$^2$ on the basis of obtained data arrays. This fixed bin size was chosen because, firstly, we wanted to make sure all the events are analyzed the same way, secondly, with this particular bin size for each event there are no less than 15 bins, and at the same time number of bins for low $j_z$-values was not so large, and the number of empty bins for high $j_z$-values was small enough. Empty bins were excluded from dataset for further analysis. After all, absolute values of the obtained values of the bin centers ($x$-data) are taken, and also the natural logarithm of these absolute values and the number of $j_z$ in each bin (counts, $y$-data) is taken. The Kolmogorov–Smirnov test for positive and negative values of bins centers was done with the use of the \texttt{kstest2} function in Matlab. The test does not reject the null hypothesis (that tested vectors of data are from the same continuous distribution) at the 1\% significance level. Therefore, we can examine absolute value $|j_z|$ instead of analysis of positive and negative $j_z$ values separately. This approach doubles our dataset that is important for the adequate fit of high $|j_z|$ values. In the end, for each event we have 30--40 bins, which is much larger than the number of free parameters of the models we use (see below). Before the fitting procedure, histogram counts are normalized to the maximum value, so that histograms should be considered as the approximation of probability density function (PDF) of $|j_z|$. To obtain PDF, counts should be normalized to the total number of datapoints in $|j_z|$ array, but we use this approach so that $y$-data varies from 0 to 1. This decision does not affect the results of this study.\\

Examples of histograms (distributions) are shown in Figure \ref{fig1:c} and \ref{fig1:d} in log-log scale for two active regions and two points in time: before and after the flare. For all the other 46 events considered histograms are similar. For low $|j_z|$-values distribution has Gaussian-like shape, for high $|j_z|$-values distribution looks like sloping "tail". Based on that form, we made approximations with three model functions:

\begin{itemize}
    \item \textbf{Model 1: Gaussian and power function.} Datapoints were divided into two groups: the first one with $n > 5$ first datapoints with the smallest $|j_z|$-value ($x$-data) ($|j_z|$ is sorted in ascending order). The second one with the rest $(N-n) > 5$ datapoints and datapoint with the highest $|j_z|$-value from the first group, so that these two groups have one common datapoint. We will refer to this datapoint as transition point or tp. The first group was fitted with Gaussian function in log-log scale, the second one -- with power function in log-log scale. $n$ (which unambiguously determines transition point) was chosen to minimize residuals ($|y_{data}-y_{model}|$). \\
    
\begin{figure}[t!]
\centering  
\subfigure[]{\label{fig1:a}\includegraphics[width=0.4\textwidth]{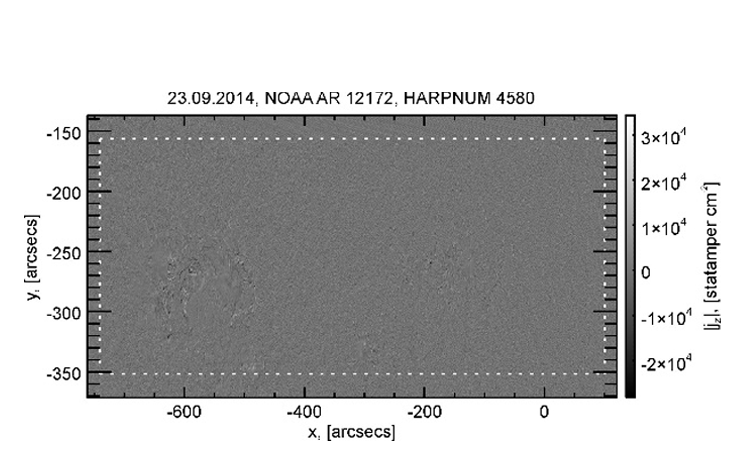}}
\subfigure[]{\label{fig1:b}\includegraphics[width=0.4\textwidth]{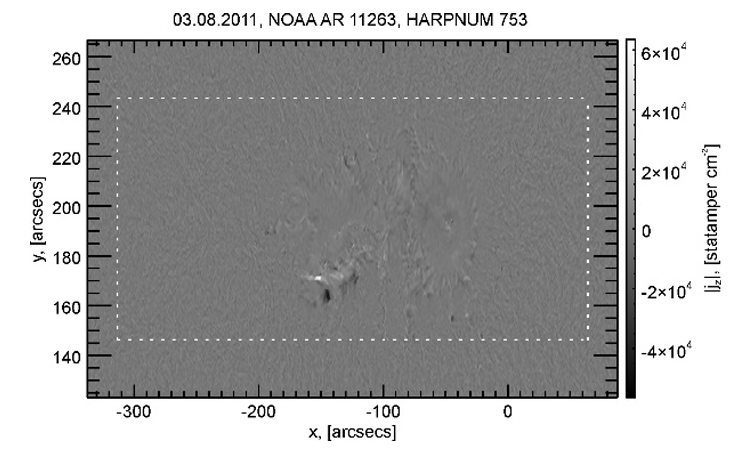}} \\
\subfigure[29.09.2014]{\label{fig1:c}\includegraphics[width=0.4\textwidth]{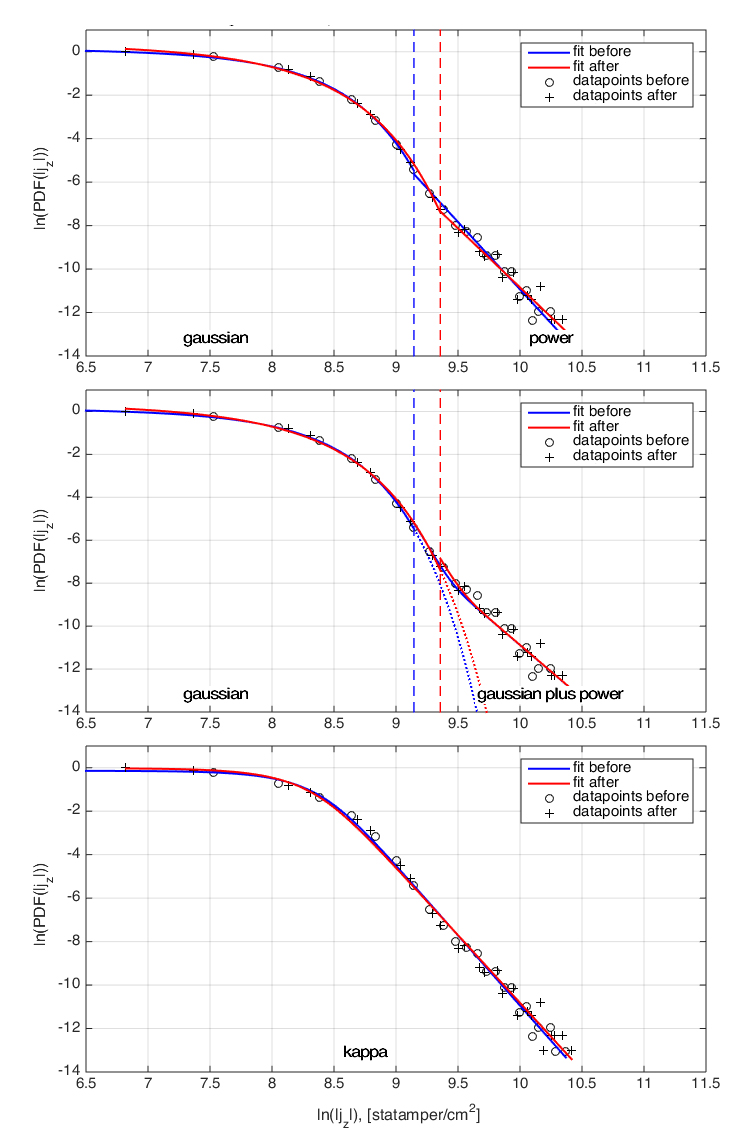}}
\subfigure[03.08.2011]{\label{fig1:d}\includegraphics[width=0.4\textwidth]{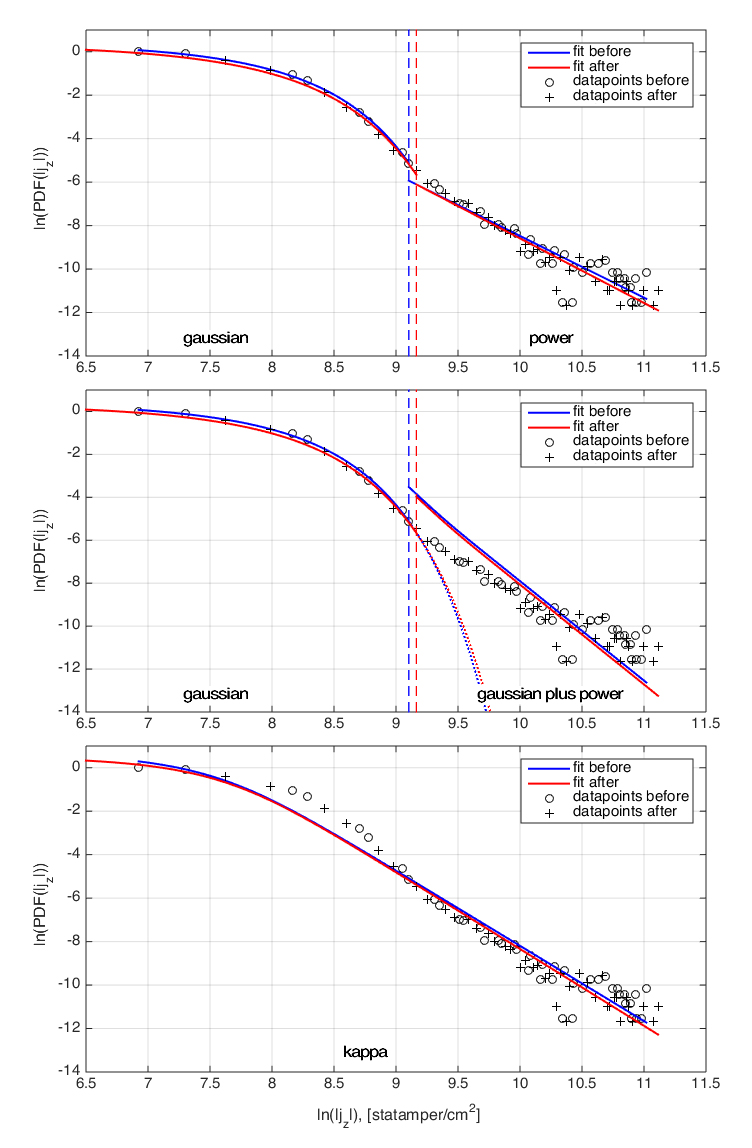}}
\caption{Examples of $|j_z|$ maps for the active regions NOAA 12172 (a) and 11263 (b) and fitted distributions of $|j_z|$ in log-log scale in (c) and (d) respectively. Distributions of $|j_z|$ before and after the flare are shown be circles and crosses respectively. Three different model fits are shown on different panels: top panel demonstrates Model 1 fit, middle panel -- Model 2 fit, and bottom panel -- Model 3 fit. Blue lines correspond to the fit of distribution before the flare, red lines correspond to the fit of distribution after the flare. Vertical dashed lines represent transition point for each dataset. Inner edge of the calculation area for "noise" distributions of $|j_z|$ is shown by the white dotted lines in (a) and (b). Corresponding "noise" distributions are shown in Figure \ref{fig7}.}
\label{fig1}
\end{figure}
    
    Model 1 in log-log scale:
    \begin{equation}
    f_{1}(x)=\left\{\begin{array}{r}{a-\left(\left[e^{x}-b\right] / c\right)^{2}, x \leq x_{t p}}, \\ {f x+d, x \geq x_{t p}}.\end{array}\right.
    \end{equation}

    Model 1 in normal scale:
    \begin{equation}
    \hat{f}_{1}(x)=\left\{\begin{aligned} A e^{-([x-b] / c)^{2}},  x \leq e^{x_{t p}}, \\ D x^{f},  x \geq e^{x_{t p}}; \end{aligned}\right.
    \end{equation}

    where $x_{tp}$ is a transition point, $A = e^a$ and $D = e^d$.
    
    \item \textbf{Model 2: Gaussian and sum of Gaussian and power function.} The same groups of datapoints as in Model 1 (with the same transition point) were fitted. The first group was fitted with Gaussian function (with the same output parameters as in Model 1) in log-log scale, the second group was fitted with sum of Gaussian (with fixed parameters obtained from the group one, so Gaussian is continuous) and power function in log-log scale.\\
    
    Model 2 in log-log scale:
    \begin{equation}
    f_{2}(x)=\left\{\begin{array}{r}{a-\left(\left[e^{x}-b\right] / c\right)^{2}, x \leq x_{t p}}, \\ {\ln \left(\exp \left[a-\left(\left[e^{x}-b\right] / c\right)^{2}\right]+d e^{f x}\right), x \geq x_{t p}.}\end{array}\right.
    \end{equation}

    Model 2 in normal scale:
    \begin{equation}
    \hat{f}_{2}(x)=\left\{\begin{aligned} A e^{-([x-b] / c)^{2}}, x \leq e^{x_{t p}}, \\ A e^{-([x-b] / c)^{2}}+D x^{f}, x \geq e^{x_{t p};} \end{aligned}\right.
    \end{equation}
    
    where $x_{tp}$ is a transition point, $A = e^a$ and $D = e^d$.
    
    \item \textbf{Model 3: kappa function.} Whole dataset was fitted with kappa function with fixed parameter $k = 0.5$.\\
    
    Model 3 in log-log scale:
    \begin{equation}
    f_{3}(x)=a+\frac{1}{k} \ln \left(\sqrt{1+k^{2} b^{2} e^{2 c x}}-k b e^{c x}\right).
    \end{equation}
    
    Model 3 in normal scale:
    \begin{equation}
    \hat{f}_{3}(x)=A(\sqrt{1+k^{2} \widetilde{x}^{2}}+k \widetilde{x})^{1 / k},
    \end{equation}
    
    where $\widetilde{x}=-b x^{c}$ and $ A=e^{a}$.

\end{itemize}

All the fits were performed with the use of \texttt{nlinfit} function in Matlab. This function uses the Levenberg-Marquardt nonlinear least squares algorithm. The adjusted coefficient of determination was calculated for each fit to determine goodness of fit:
\begin{equation}
R_{adj}^{2}=1-\frac{SS_{res} /(n-k)}{SS_{tot} /(n-1)},
\end{equation}

where $SS_{res}$ is the residual sum of squares, $SS_{tot}$ is the total sum of squares, $n$ is the sample size and $k$ is the number of parameters in a model. The closer $R_{adj}^{2}$ is to 1, the better fit is. For Models 1 and 2 mean (between the two groups of datapoints) $R_{adj}^{2}$ was calculated.\\

After we had obtained all model parameters, we checked whether there is correlation between them and X-ray flare class (determined by the Geostationary Operational Environmental Satellite -- GOES) of corresponding flares, and between them and Hale magnetic class of the parent active regions (the Mount Wilson classification). This information was taken from the SolarMonitor website (https://solarmonitor.org/).

\begin{figure}[h]
\centering  
\subfigure[]{\label{fig2:a}\includegraphics[width=0.32\textwidth]{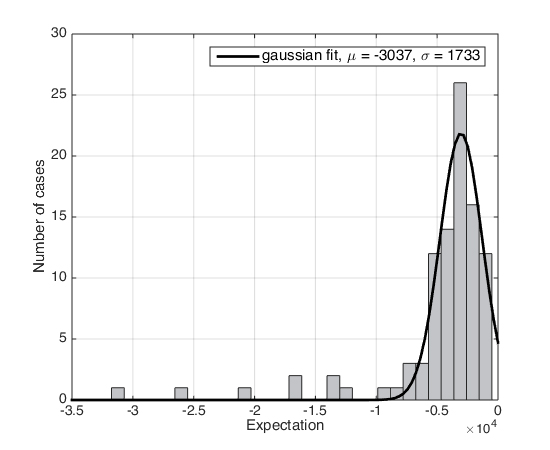}}
\subfigure[]{\label{fig2:b}\includegraphics[width=0.32\textwidth]{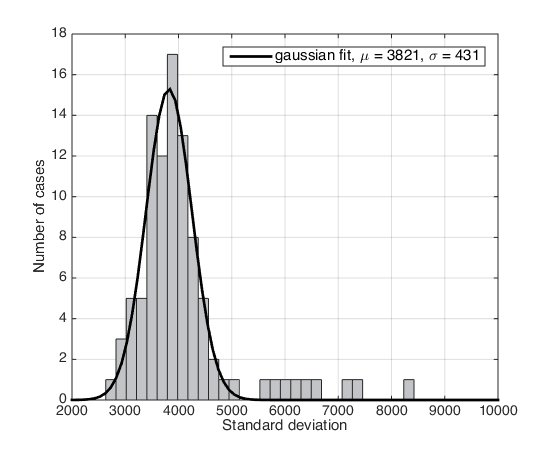}} 
\subfigure[]{\label{fig2:c}\includegraphics[width=0.32\textwidth]{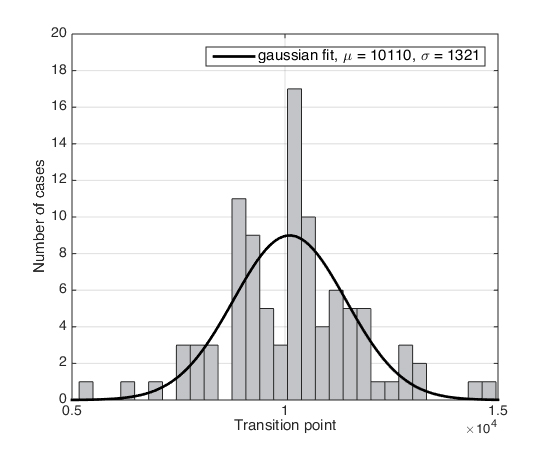}}\\
\subfigure[]{\label{fig2:d}\includegraphics[width=0.32\textwidth]{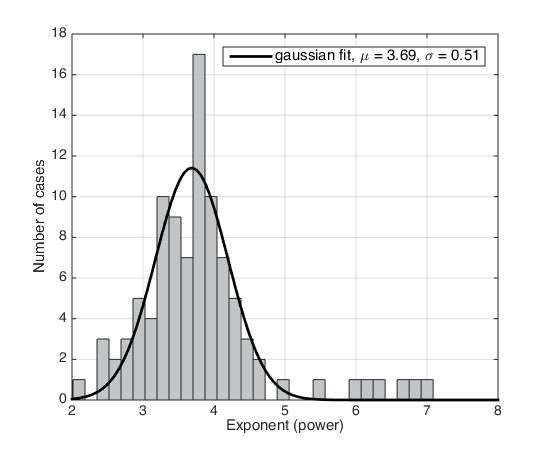}}
\subfigure[]{\label{fig2:e}\includegraphics[width=0.32\textwidth]{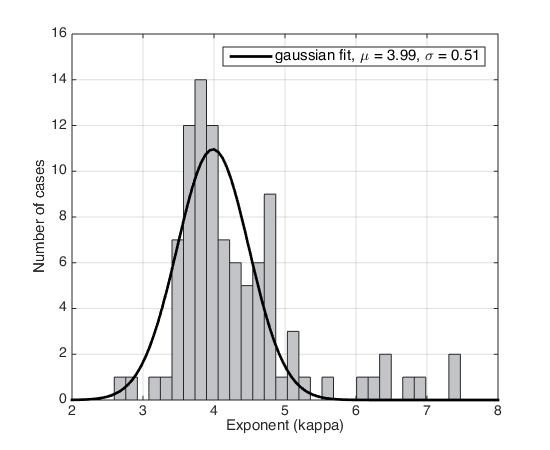}}
\caption{Histograms of Model 1 and Model 3 parameters obtained by least square fit of 96 $|j_z|$ distributions (48 before the flares and 48 after the flares). (a): expectation of Gaussian $\mu = b$, Model 1; (b): standard deviation of Gaussian $\sigma = c/\sqrt{2}$, Model 1; (c): transition point $x$-value $|j_z|_{tp}$, Model 1; (d): absolute value of power function exponent $|f|$, Model 1; (e): exponent of kappa function $c/k$, Model 3. Result of Gaussian fit of histogram data is shown by the solid black curve. Standard deviation $\sigma$ and expectation $\mu$ for each parameter based on Gaussian fit of histograms are shown in the legend of each panel.}
\label{fig2}
\end{figure}

\section{Results}

From total 96 distributions of $|j_z|$ (48 before the flares and 48 after the flares) only 34 (35\%) of them have $R_{adj}^{2} < 0.95$ according to Model 1, 74 (77\%) of them according to Model 2, and 10 (10\%) of them according to Model 3. Moreover, visual examination of the fit results showed that Model 1 adequately approximates data in all the cases, while Model 2 approximates data much worse. Model 3 gives adequate approximation in most cases, especially for high $|j_z|$-values. In Figure \ref{fig1} two examples are shown. One can see that for the active region 12172 (Figure \ref{fig1:c}) all three Models agree with data, but for the active region 11263 (Figure \ref{fig1:d}) only Model 1 agrees with data. In this case Model 2 does not match with the "tail" of distribution and Model 3 does not match with Gaussian-like part of distribution for low $|j_z|$-values. Based on that, we will only consider Models 1 and 3 in further discussion. It is worth mentioning that parameter $k = 0.5$ was fixed in Model 3. As the experiment, we made similar fit for several active regions, but with non-fixed $k$. As the result, $R_{adj}^{2}$ did not change much and neither did the exponent of the function $c/k$. By this reason, we consider only the fit results of Model 3 with fixed $k = 0.5$. \\

The histograms were made for Model 1 parameters: expectation of Gaussian $\mu = b$; standard deviation of Gaussian $\sigma = c/\sqrt{2}$; transition point between Gaussian and power function $|j_z|_{tp}$; absolute value of power function exponent $|f|$. For Model 3 we present only the histogram of the absolute value of kappa function exponent $c/k$. We consider these Model parameters to be potentially the most important ones and do not take normalization constants into account. Each histogram was later fitted with Gaussian function to obtain expectation $\mu$ and standard deviation $\sigma$ of each parameter considered. Histograms with Gaussian fits are shown in Figure \ref{fig2}. The obtained $\mu$ and $\sigma$ are shown in each panels legend and in Table \ref{tab1}.

\begin{table}[h]
    \centering
    \begin{tabular}{c|c|c}
Model & Parameter & Value\\ \hline
1 & Expectation of Gaussian & $\mu = -3037 \pm 1733$ statampere/cm$^2$ \\
1 & Standard deviation of Gaussian & $\sigma = 3821 \pm 431$ statampere/cm$^2$ \\
1 & Transition point & $|j_z|_{tp} = 10110 \pm 1321$ statampere/cm$^2$ \\
1 & Absolute value of power function exponent & $|f| = 3.69 \pm 0.51$\\
3 & Absolute value of kappa function exponent & $c/k = 3.99 \pm 0.51$
    \end{tabular}
    \caption{Model 1 and 3 parameters obtained from Gaussian fit of histograms shown in Figure \ref{fig2}.}
    \label{tab1}
\end{table}

Firstly, the expectation of $|j_z|$ distribution turned out to be negative, which is simply the fitting result not related to the physics. We tried to do fitting with fixed $\mu = 0$ and goodness of fit almost did not change. Secondly, one can see that Model parameters can be described by Gaussian distribution and their values are not widely dispersed from the expectation. Thirdly, the exponents of power and kappa functions have close values. This is the argument that the "tail" of $|j_z|$ distribution can be described by both Models and is a power-law "tail".\\

\begin{figure}[b!]
\centering  
\subfigure[]{\label{fig3:a}\includegraphics[width=0.32\textwidth]{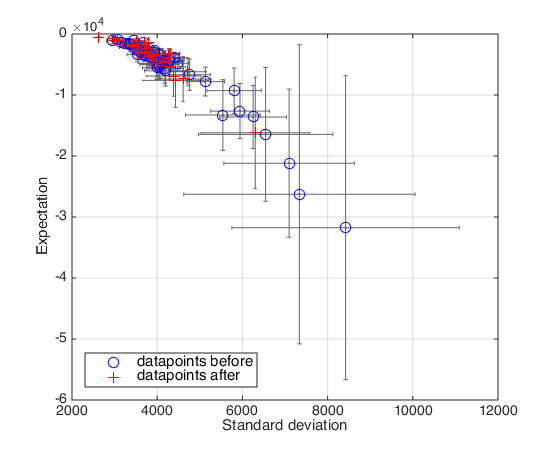}}
\subfigure[]{\label{fig3:b}\includegraphics[width=0.32\textwidth]{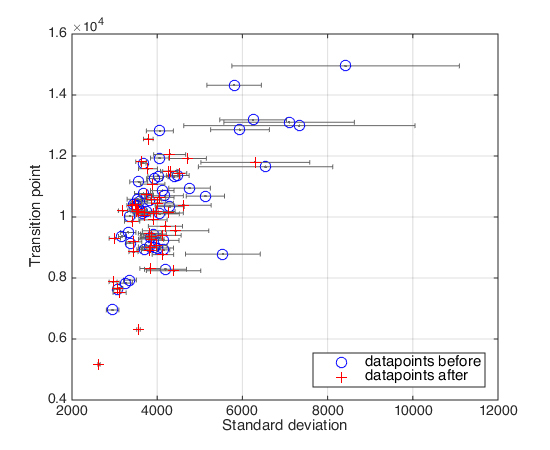}} 
\subfigure[]{\label{fig3:c}\includegraphics[width=0.32\textwidth]{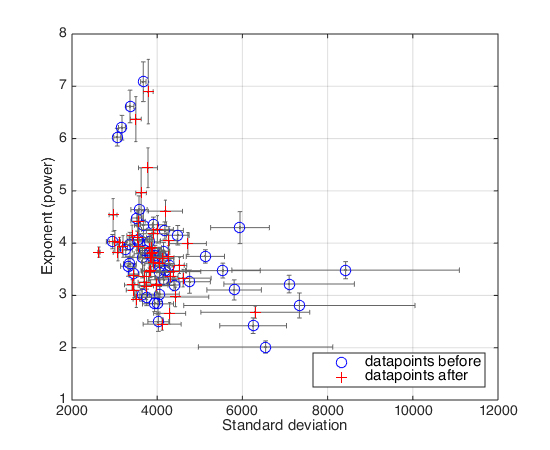}}\\
\subfigure[]{\label{fig3:d}\includegraphics[width=0.32\textwidth]{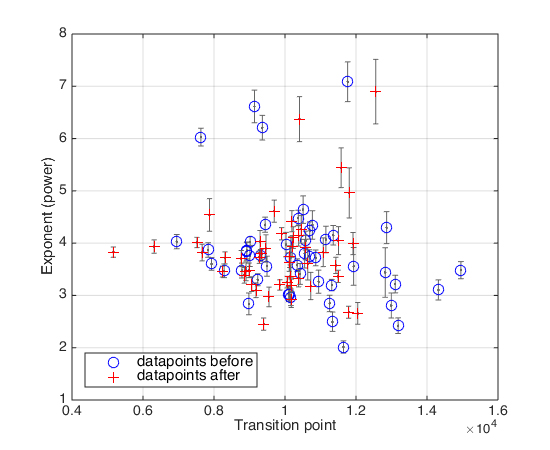}}
\subfigure[]{\label{fig3:e}\includegraphics[width=0.32\textwidth]{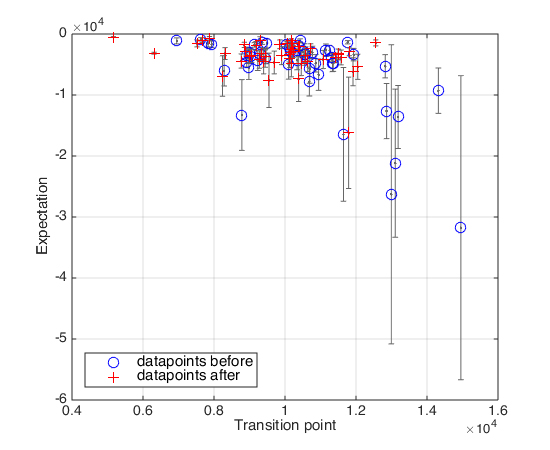}}
\subfigure[]{\label{fig3:f}\includegraphics[width=0.32\textwidth]{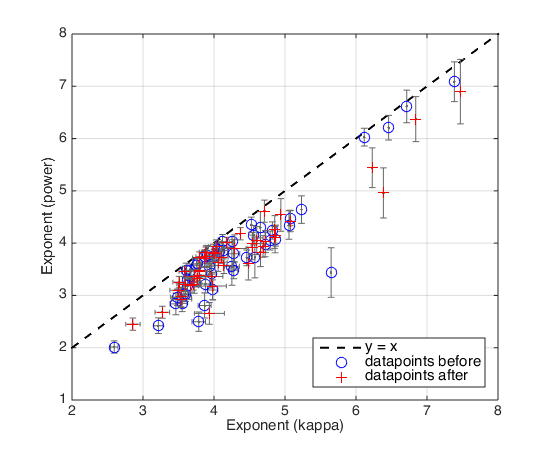}}
\caption{Mutual scalings of parameters for 96 distributions of $|j_z|$. Datapoints corresponding to the times before (after) the flares are shown by blue circles (red crosses). Errors of the parameters are shown by solid grey segments. (a) Gaussian expectation $\mu$ is plotted against Gaussian standard deviation $\sigma$; (b) $|j_z|$-value of transition point $x_{tp}$ is plotted against Gaussian standard deviation $\sigma$; (c) absolute value of the power function exponent $|f|$ is plotted against Gaussian standard deviation $\sigma$; (d) absolute value of the power function exponent $|f|$ is plotted against $|j_z|$-value of transition point $x_{tp}$; (e) Gaussian expectation $\mu$ is plotted against $|j_z|$-value of transition point $|j_z|_{tp}$; (f) absolute value of the power function exponent $|f|$ is plotted against the kappa function exponent value $c/k$. $x = y$ is shown by the black dashed line.}
\label{fig3}
\end{figure}

For the examination of possible relationship between these parameters several mutual scaling plots were done. They are shown in Figure \ref{fig3}. The corresponding Pearson correlation coefficients were calculated and presented with 95\% confidence bounds in Table \ref{tab2}.\\

\begin{table}[h]
    \centering
    \small
    \begin{tabular}{c | c|c|c|c|c}
 & $\sigma$ & $\mu$ & $|j_z|_{tp}$ & $|f|$ & $c/k$\\ \hline
$\sigma$ & \cellcolor[HTML]{d2d7d3} & -0.74 [-0.82 ;-0.63] & +0.61 [0.46; 0.72] & -0.36 [-0.53; -0.18] & - \\ \hline
$\mu$ & -0.74 [-0.82 ;-0.63] & \cellcolor[HTML]{d2d7d3} & -0.35 [-0.51; -0.16] & - & - \\ \hline
$|j_z|_{tp}$ & +0.61 [0.46; 0.72] & -0.35 [-0.51; -0.16] & \cellcolor[HTML]{d2d7d3} & -0.08 [-0.28; 0.12] & - \\ \hline
$|f|$ & -0.36 [-0.53; -0.18] & - & -0.08 [-0.28; 0.12] &\cellcolor[HTML]{d2d7d3} & 0.92 [0.88; 0.95] \\ \hline
$c/k$ & - & - & - & 0.92 [0.88; 0.95] & \cellcolor[HTML]{d2d7d3}
    \end{tabular}
    \caption{Mutual correlation coefficients for five parameters: Gaussian standard deviation $\sigma$ (Model 1), Gaussian expectation $\mu$ (Model 1), transition point between Gaussian and power function $|j_z|_{tp}$ (Model 1), absolute value of power function exponent $|f|$ (Model 1), kappa function exponent $c/k$ (Model 3). Numbers in square brackets correspond to 95\% confidence bounds.}
    \label{tab2}
\end{table}

\normalsize

\begin{figure}[b!]
\centering  
\subfigure[]{\label{fig4:a}\includegraphics[width=0.32\textwidth]{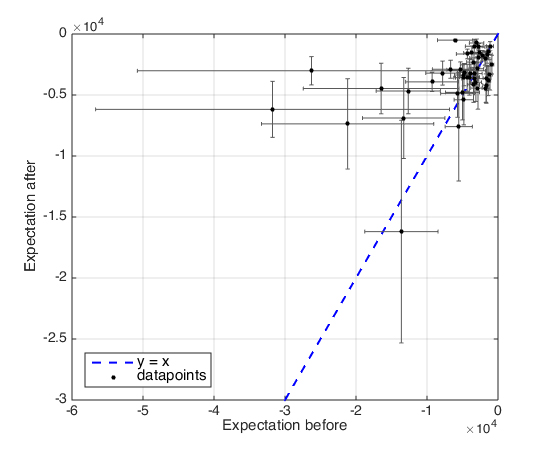}}
\subfigure[]{\label{fig4:b}\includegraphics[width=0.32\textwidth]{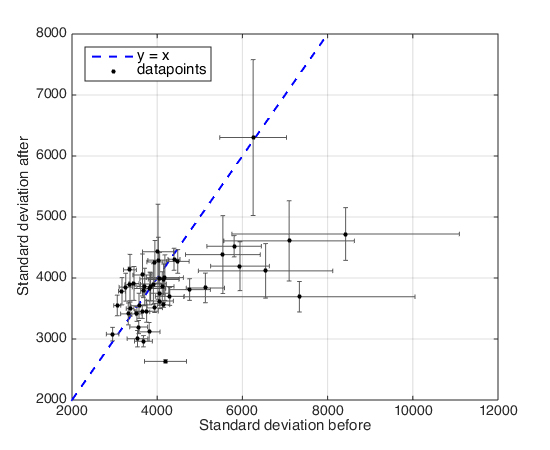}} 
\subfigure[]{\label{fig4:c}\includegraphics[width=0.32\textwidth]{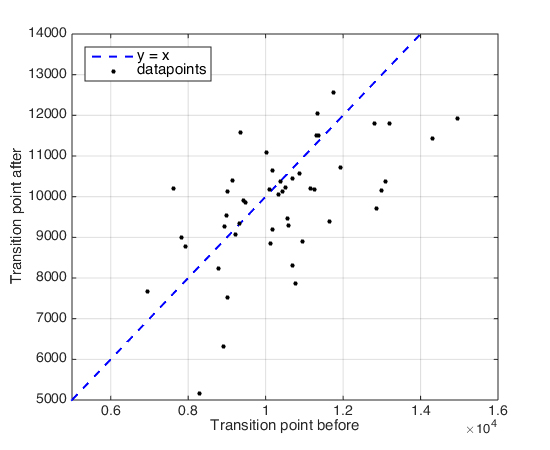}}\\
\subfigure[]{\label{fig4:d}\includegraphics[width=0.32\textwidth]{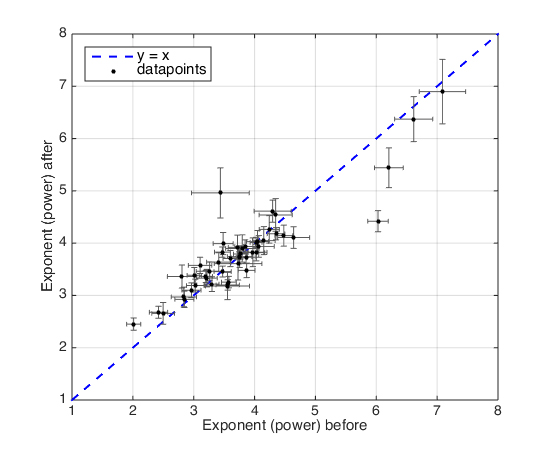}}
\subfigure[]{\label{fig4:e}\includegraphics[width=0.32\textwidth]{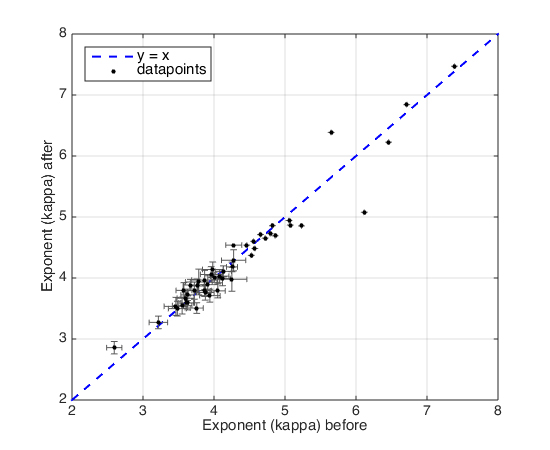}}
\caption{Parameters of Models of $|j_z|$ distributions for 48 active regions before (x-axis) and after (y-axis) the flares. Errors of the parameters are shown by the thin solid grey segments. The $x = y$ line is shown by the thick dashed blue line. (a) Gaussian expectation $\mu = b$, Model 1; (b) Gaussian standard deviation $\sigma = c/\sqrt{2}$, Model 1; (c) transition point $|j_z|_{tp}$, Model 1; (d) absolute value of the power function exponent $|f|$, Model 1; (e) kappa function exponent $c/k$, Model 3.}
\label{fig4}
\end{figure}

A high correlation coefficient between the kappa function exponent of Model 3 and the power function exponent of Model 1 can be noticed. As it was stated before, this indicates that the "tail" of $|j_z|$ distribution is power-law, but the kappa function exponent is a bit higher than the power function exponent in general. Also one can note correlation between the standard deviation of Gaussian and the transition point. This connection suggests that the wider Gaussian distribution is, the higher $|j_z|$-value for which the power-law "tail" stands out (see Discussion section).  \\

For all the 48 active region the above parameters of Models 1 and 3 were compared for times before and after the flares (Figure \ref{fig4}) and corresponding Pearson correlation coefficients were calculated. Figure \ref{fig4:a}: Gaussian expectation $\mu$, 0.45 [0.19; 0.65]; Figure \ref{fig4:b}: Gaussian standard deviation $\sigma$, 0.55 [0.32; 0.72]; Figure \ref{fig4:c}: transition point $|j_z|_{tp}$, 0.56 [0.33; 0.73]; Figure \ref{fig4:d}: absolute value of power function exponent $|f|$, 0.91 [0.84; 0.95]; Figure \ref{fig4:e}: kappa function exponent $c/k$, 0.97 [0.94; 0.98]. The numbers in square brackets correspond to 95\% confidence bounds. The strongest correlations are found for the power function exponent (Model 1) and for kappa function exponent (Model 3). During the flare values of this parameters do not change much, in general. In Figure \ref{fig4:b} we can pick the group of dots with high values before the flares under the $y = x$ line. This is due to the fact that the expectation of the Gaussian before these flares has a higher negative value than after the flares. This fact is more likely because of the fit quality, rather than of physical reasons. Also one can notice higher errors for that type of events.\\

\begin{figure}[t!]
\centering  
\subfigure[]{\label{fig5:a}\includegraphics[width=0.32\textwidth]{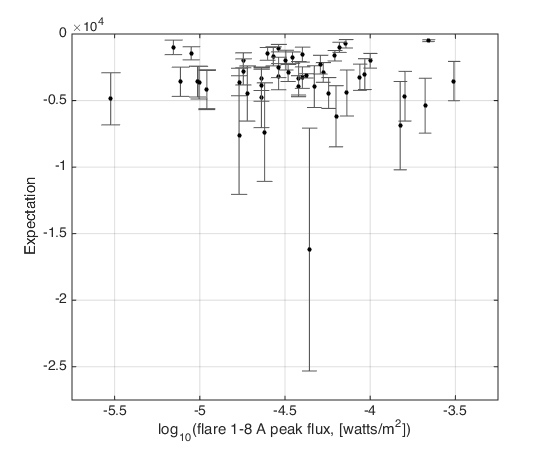}}
\subfigure[]{\label{fig5:b}\includegraphics[width=0.32\textwidth]{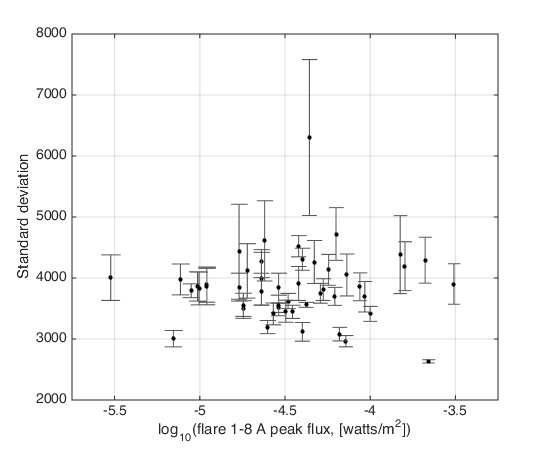}} 
\subfigure[]{\label{fig5:c}\includegraphics[width=0.32\textwidth]{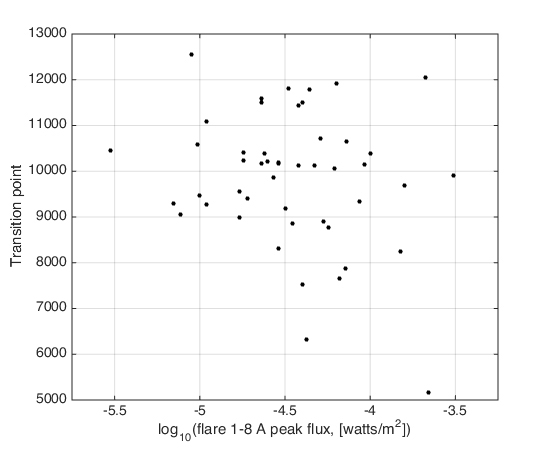}}\\
\subfigure[]{\label{fig5:d}\includegraphics[width=0.32\textwidth]{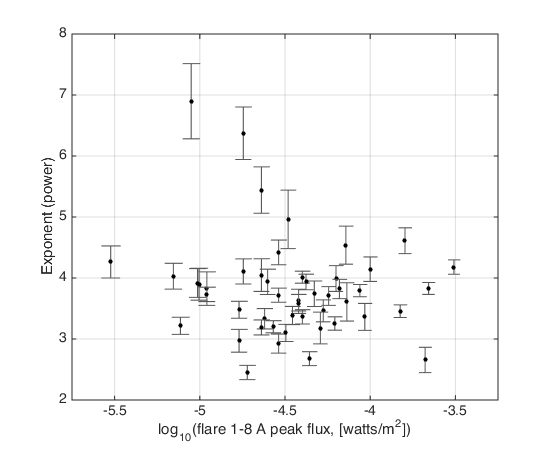}}
\subfigure[]{\label{fig5:e}\includegraphics[width=0.32\textwidth]{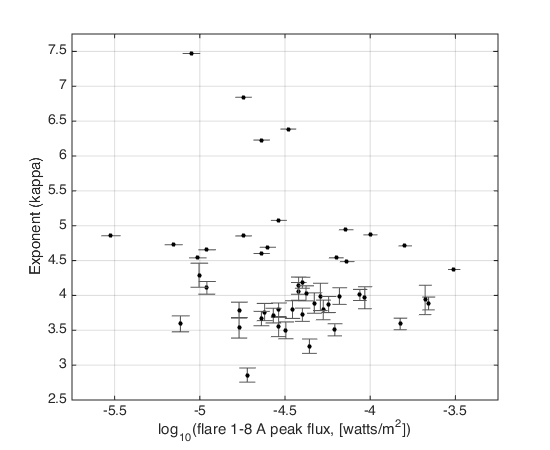}}
\caption{Parameters of Models 1 and 3 of pre-flare $|j_z|$ distributions plotted against the decimal logarithm of the flare peak flux in 1-8 \AA \ GOES channel. Errors of the parameters are shown by the thin solid grey segments. (a) Gaussian expectation $\mu = b$, Model 1; (b) Gaussian standard deviation $\sigma = c/\sqrt{2}$, Model 1; (c) transition point $|j_z|_{tp}$, Model 1; (d) absolute value of the power function exponent $|f|$, in Model 1; (e) kappa function exponent $c/k$, Model 3.}
\label{fig5}
\end{figure}

\begin{figure}[h]
\centering  
\subfigure[]{\label{fig6:a}\includegraphics[width=0.32\textwidth]{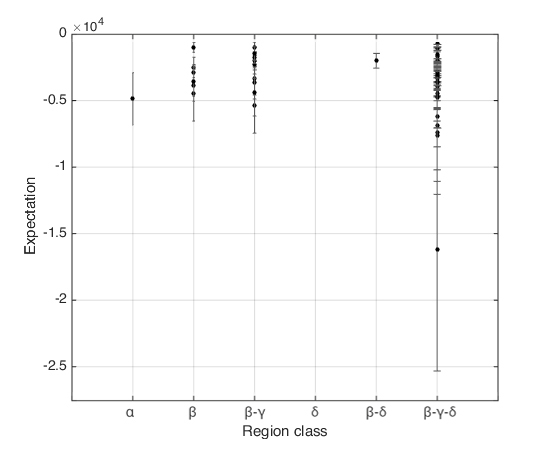}}
\subfigure[]{\label{fig6:b}\includegraphics[width=0.32\textwidth]{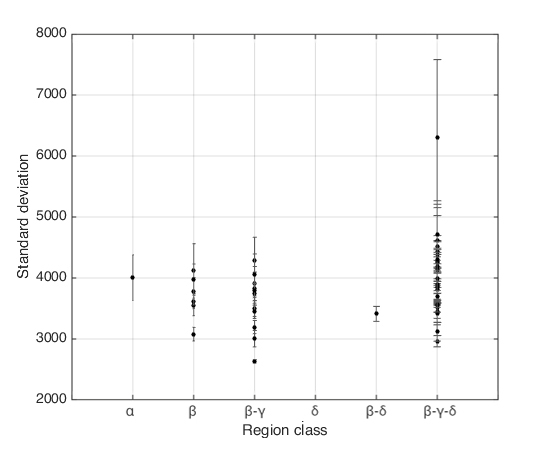}} 
\subfigure[]{\label{fig6:c}\includegraphics[width=0.32\textwidth]{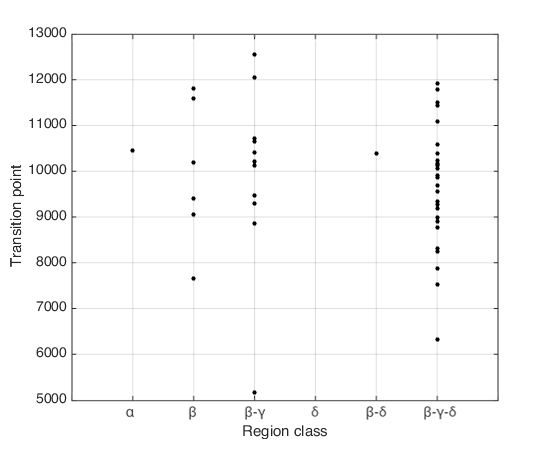}}\\
\subfigure[]{\label{fig6:d}\includegraphics[width=0.32\textwidth]{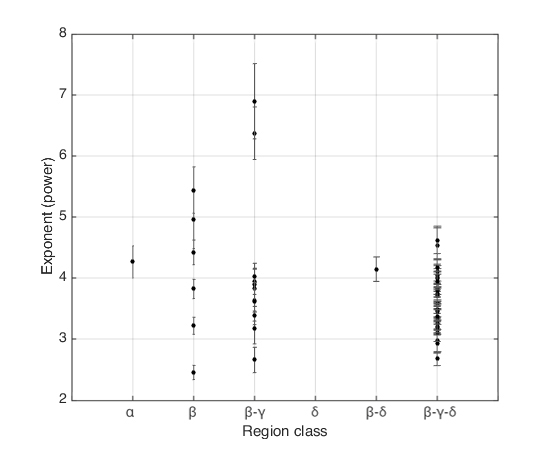}}
\subfigure[]{\label{fig6:e}\includegraphics[width=0.32\textwidth]{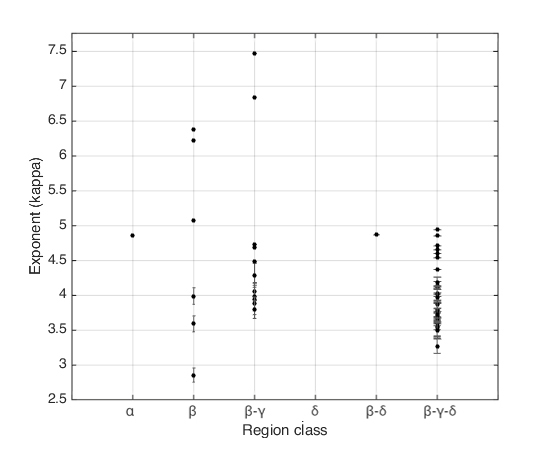}}
\caption{Parameters of Models 1 and 3 of pre-flare $|j_z|$ distributions plotted against Hale magnetic class of 48 active regions considered. Errors of the parameters are shown by the thin solid grey segments. (a) Gaussian expectation $\mu = b$, Model 1; (b) Gaussian standard deviation $\sigma = c/\sqrt{2}$, Model 1; (c) transition point $|j_z|_{tp}$, Model 1; (d) absolute value of the power function exponent $|f|$, Model 1; (e) kappa function exponent $c/k$, Model 3.}
\label{fig6}
\end{figure}

Additionally, we checked whether there is a correlation between Model parameters and both X-ray flare class and Hale magnetic class of the active regions. In Figure \ref{fig5} parameters are plotted against the common logarithm of the flare peak flux in 1-8 \AA \ GOES channel. In Figure \ref{fig6} parameters are plotted against the Hale magnetic class of the active regions. One can see no obvious dependencies. The one thing we only point out is that most of the active regions considered in this study (29 or 60\%) have $\beta\gamma\delta$ class. This seems quite natural, since the selected flares were, in general, quite powerful and were accompanied by hard X-ray emission ($>50$ keV), and it is known that $\beta\gamma\delta$ class regions tend to produce more flares, including powerful ones \cite{2019LRSP...16....3T}.

\section{Discussion}

The most important result of this study is establishing the form of distribution (PDF) of absolute value of photospheric vertical electric current density $|j_z|$, calculated based on vector magnetograms SHARP\_CEA obtained from HMI/SDO for 48 active regions with flare activity. The best-describing Model (from all three considered) is Gaussian for low $|j_z|$-values and decreasing power function for higher $|j_z|$-values. The transition point between the Gaussian and the power-law "tail" has an average value of $|j_z|_{tp} = 10110 \pm 1321$ statampere/cm$^2$. One can notice that the form of PDF($|j_z|$) is similar to the shape of an X-ray energy spectrum in solar flares in the range of $\approx 1-100$ keV \cite{2002SoPh..210....3L}. However, we think that this is just a coincidence. We suggest that the Gaussian form of PDF($|j_z|$) for low $|j_z|$-values is determined by the noise of the vector magnetograms used, while the power-law "tail" can be due to the physics of magnetic fields and electric currents in solar active regions.\\

\begin{figure}[t]
\centering  
\subfigure[]{\label{fig7:a}\includegraphics[width=0.4\textwidth]{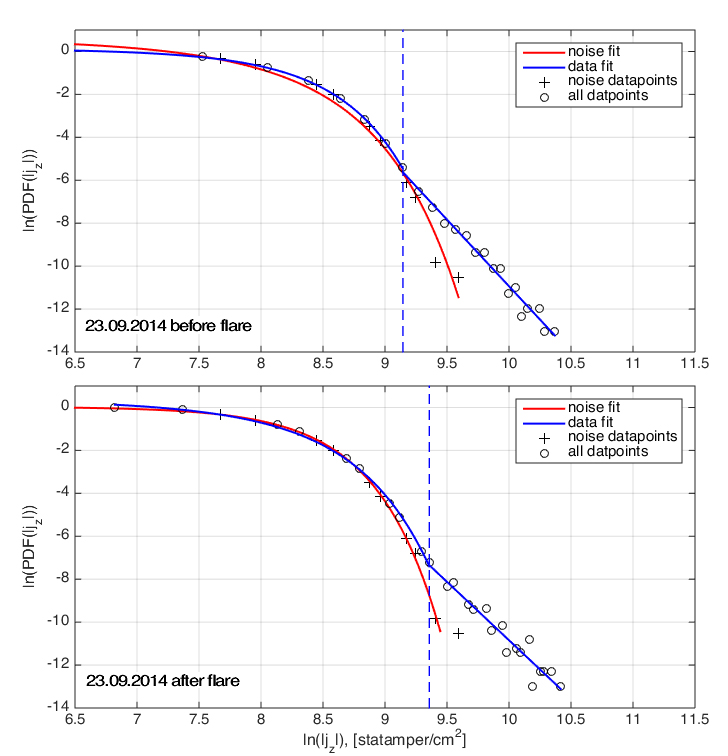}}
\subfigure[]{\label{fig7:b}\includegraphics[width=0.4\textwidth]{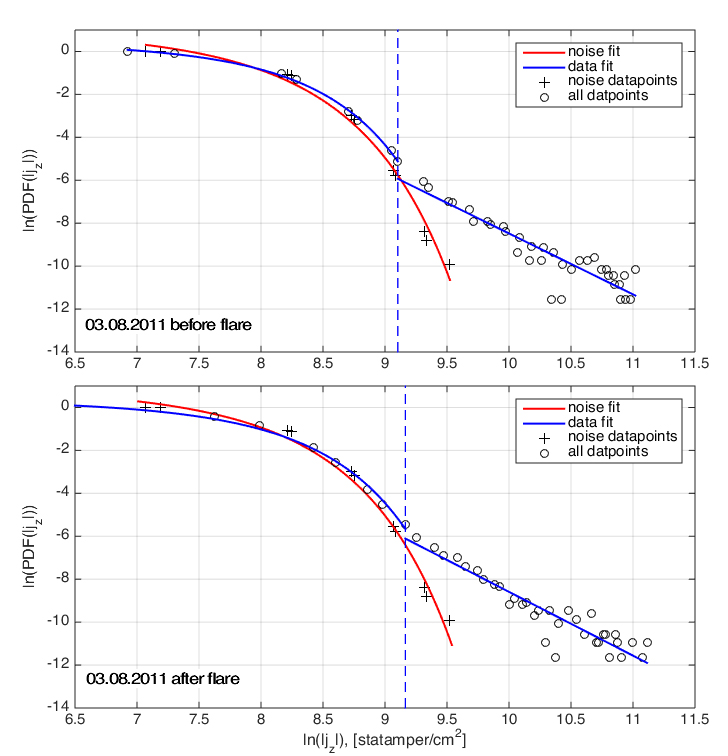}} 
\caption{Distributions of $|j_z|$ before the flare (top panels) and after the flare (bottom panels) for the active regions 12172 (a) and 11263 (b). Datapoints obtained from the whole SHARP active region are shown by circles and datapoints obtained from only the "noise" edges are shown by crosses. Least square fit of the "noise" data is shown by the solid red curve, fit of the whole dataset is shown by the blue solid curves. Blue dashed vertical line corresponds to the transition point. }
\label{fig7}
\end{figure}

To justify the assumption about the instrumental (noise) nature of the Gaussian form of distribution at low $|j_z|$-values, we compared the distribution of $|j_z|$ for the whole area of an active region defined in SHARP, with a distribution calculated only for the edges of an active region. There is no strong fields and $|j_z|$ at the edges of active regions, consequently we can consider them as the background ("noise") regions of the HMI/SDO magnetograms. Strips with a width of 50 pixels around the perimeter of an active region were examined. In Figure \ref{fig1:a} and \ref{fig1:b} one can see these edges shown by the dotted white lines for two active regions: 12172 and 11263. Distributions of $|j_z|$ for all the area and for only the edges are shown in Figure \ref{fig7}. It can be seen that the distribution of $|j_z|$ for the "noise" region has form of Gaussian, while distribution for the entire active region has the form of Gaussian for low values and a power-law “tail” for higher values. Moreover, the Gaussian for the "noise" region is very close to the Gaussian for the whole region. \\

As the additional proof to the noise nature of Gaussian, we can estimate the error of the component of the magnetic field transverse to the line-of-sight $\sigma(B_\perp)$ from the obtained standard deviation of Gaussian fit of $|j_z|$ distribution, $\sigma(|j_z|) = 3821 \pm 431$ statampere/cm$^2$. We obtain $\sigma(B_\perp) \approx \left[4 \pi \Delta l \sigma(|j_{z}|)\right] /[c \sqrt{2}] \approx 41$ G, where $\Delta l \approx 3.6 \cdot 10^7$ cm is the linear size of HMI/SDO pixel on the photosphere and $c$ is the speed of light in vacuum. This resulting value of $\sigma(B_\perp)$ lies between the boundary values of 20 G (before 2014) and 50 G (after 2014), determined for the transverse component of magnetic field for the HMI/SDO vector magnetograms (DOFFSET parameter, \cite{2014SoPh..289.3483H}). This is a strong argument in favor of the assumption made. \\ 

The fact that the transition point between the Gaussian and the power-law "tail" of $|j_z|$-distribution is observed at $|j_z|_{tp} = 10110 \pm 1321$, which is similar to the tripled Gaussian standard deviation $3\sigma_{stdev} = 11463 \pm 1293$ statamper/cm$^2$, is an argument in favor that the power-law "tail" of $|j_z|$-distribution is not a data noise. This indicates that when studying vertical currents on the photosphere using HMI/SDO data, the "three sigma" rule should be used and only values greater than $3\sigma_{stdev}$ should be considered, but lower $|j_z|$-values should be treated with caution.\\

The presence of the power-law "tail" in the $|j_z|$-distributions in active regions of the Sun is an interesting fact. It can indicate specific turbulent nature of the electric currents formation/dissipation processes. In essence, this is not surprising, since it is known that the distributions of various characteristics of the photospheric magnetic fields, in particular, magnetic flux \cite{2005ApJ...619.1160A}, from which $j_z$ is obtained by derivation, have a power-law form. The magnetic power spectrum also has a power-law form \cite{2005ApJ...629.1141A}, \cite{2010ApJ...720..717A}. The power-law form is also inherent in the spatial characteristics of the current helicity in active regions \cite{2018MNRAS.480.3780K}. We found that for the 48 active regions considered, the absolute value of the power function exponent of $|j_z|$-distribution is concentrated in the vicinity of 3.69 with a small standard deviation of 0.51. For the kappa function, similar values ($3.99 \pm 0.51$) are obtained. The question of what physical processes determine these values requires further study.\\ 

In conclusion, we note that we did not find an explicit correlation between the parameters of the considered models of PDF($|j_z|$) in active regions studied and X-ray (GOES) class of the flares happened there. This can be interpreted by the fact that the model parameters were determined by the distribution of an entire active region with linear scales of several hundred arc seconds, while a flare is a local process that usually occupies a small part of a parent active region (several arc seconds or tens of arc seconds). In the future, it seems interesting to statistically study the relationship between the characteristics of flares and the parameters of local $|j_z|$-distributions in flare regions, in particular, in the vicinity of photospheric magnetic polarity inversion lines, where flares usually occur. The absence of a connection between the $|j_z|$-distribution parameters and the Hale magnetic class of the active regions can be explained by insufficient statistics or the excessively descriptive (non-quantitative) nature of the Hale classification of active regions, without diminishing its merits.

\section{Acknowledgments}

We are grateful to the Helioseismic and Magnetic Imager on board the Solar Dynamics Observatory team for the SHARP series vector magnetograms freely available online. Some data supplied courtesy of SolarMonitor.org. The initial phase of this study was carried out as part of the CHINESE ACADEMY OF SCIENCES President’s International Fellowship Initiative (GRANT No. 2018VMB0007) project. Most of the work (data processing and analysis, preparation of the article) was supported by the Russian Science Foundation grant (project No. 17-72-20134).

\printbibliography 

@PROCEEDINGS{1994ASSL..189.....S,
    title = "{Solar Magnetic Fields: Polarized Radiation Diagnostics}",
booktitle = {Astrophysics and Space Science Library},
     year = 1994,
   series = {Astrophysics and Space Science Library},
   volume = 189,
   editor = {{Stenflo}, J.},
    month = mar,
      doi = {10.1007/978-94-015-8246-9},
   adsurl = {https://ui.adsabs.harvard.edu/abs/1994ASSL..189.....S}
}

@PROCEEDINGS{2015ASSL..417.....R,
    title = "{Physics of Magnetic Flux Tubes}",
booktitle = {Astrophysics and Space Science Library},
     year = 2015,
   series = {Astrophysics and Space Science Library},
   volume = 417,
   editor = {{Ryutova}, M.},
      doi = {10.1007/978-3-662-45243-1},
   adsurl = {https://ui.adsabs.harvard.edu/abs/2015ASSL..417.....R}
}

@BOOK{severny,
  TITLE = {Nekotorie problemi fiziki Solnca},
  AUTHOR = {{Severny}, A.~B.},
  YEAR = {1988},
  PUBLISHER = {Moskva, Nauka},
}

@INPROCEEDINGS{2018GMS...235...43F,
   author = {{Fleishman}, G.~D. and {Pevtsov}, A.~A.},
    title = "{Electric Currents in the Solar Atmosphere}",
booktitle = {Electric Currents in Geospace and Beyond},
     year = 2018,
   series = {Washington DC American Geophysical Union Geophysical Monograph Series},
   volume = 235,
   editor = {{Keiling}, A. and {Marghitu}, O. and {Wheatland}, M.},
    month = mar,
    pages = {43-65},
      doi = {10.1002/9781119324522.ch3},
   adsurl = {https://ui.adsabs.harvard.edu/abs/2018GMS...235...43F}
}

@ARTICLE{2010ApJ...721L..58P,
   author = {{Puschmann}, K.~G. and {Ruiz Cobo}, B. and {Mart{\'{\i}}nez Pillet}, V.
	},
    title = "{The Electrical Current Density Vector in the Inner Penumbra of a Sunspot}",
  journal = {The Astrophysical Journal Letters},
archivePrefix = "arXiv",
   eprint = {1008.2131},
 primaryClass = "astro-ph.SR",
     year = 2010,
    month = sep,
   volume = 721,
    pages = {L58-L61},
      doi = {10.1088/2041-8205/721/1/L58},
   adsurl = {https://ui.adsabs.harvard.edu/abs/2010ApJ...721L..58P}
}

@ARTICLE{2017Ap.....60..544F,
   author = {{Fursyak}, Y.~A. and {Abramenko}, V.~I.},
    title = "{Possibilities for Estimating Horizontal Electrical Currents in Active Regions on the Sun}",
  journal = {Astrophysics},
     year = 2017,
    month = dec,
   volume = 60,
    pages = {544-552},
      doi = {10.1007/s10511-017-9505-6},
   adsurl = {https://ui.adsabs.harvard.edu/abs/2017Ap.....60..544F}
}

@INPROCEEDINGS{2018GMS...235..391S,
   author = {{Schmieder}, B. and {Aulanier}, G.},
    title = "{Solar Active Region Electric Currents Before and During Eruptive Flares}",
booktitle = {Electric Currents in Geospace and Beyond},
     year = 2018,
   series = {Washington DC American Geophysical Union Geophysical Monograph Series},
   volume = 235,
archivePrefix = "arXiv",
   eprint = {1903.04050},
 primaryClass = "astro-ph.SR",
   editor = {{Keiling}, A. and {Marghitu}, O. and {Wheatland}, M.},
    month = feb,
    pages = {391-406},
      doi = {10.1002/9781119324522.ch23},
   adsurl = {https://ui.adsabs.harvard.edu/abs/2018GMS...235..391S}
}

@BOOK{zaitsevstepanov,
  TITLE = {Magnitosferi aktivnih oblastei Solnca i zvezd},
  AUTHOR = {{Stepanov}, A.~V. and {Zaitsev}, V.~V.},
  YEAR = {2019},
  PUBLISHER = {Moskva, Fizmatlit},
}

@ARTICLE{1990IzKry..81....8A,
   author = {{Abramenko}, V.~I. and {Gopasiuk}, S.~I. and {Ogir'}, M.~B.},
    title = "{Identification of varieties of solar flares based on an investigation of electric currents}",
  journal = {Izvestiya Ordena Trudovogo Krasnogo Znameni Krymskoj Astrofizicheskoj Observatorii},
     year = 1990,
   volume = 81,
    pages = {8-13},
   adsurl = {https://ui.adsabs.harvard.edu/abs/1990IzKry..81....8A}
}

@ARTICLE{2017SoPh..292..159K,
   author = {{Kontogiannis}, I. and {Georgoulis}, M.~K. and {Park}, S.-H. and 
	{Guerra}, J.~A.},
    title = "{Non-neutralized Electric Currents in Solar Active Regions and Flare Productivity}",
  journal = {Solar Physics},
archivePrefix = "arXiv",
   eprint = {1708.07087},
 primaryClass = "astro-ph.SR",
     year = 2017,
    month = nov,
   volume = 292,
      eid = {159},
    pages = {159},
      doi = {10.1007/s11207-017-1185-1},
   adsurl = {https://ui.adsabs.harvard.edu/abs/2017SoPh..292..159K}
}

@ARTICLE{2018Ge&Ae..58.1129F,
   author = {{Fursyak}, Y.~A.},
    title = "{Vertical Electric Currents in Active Regions: Calculation Methods and Relation to the Flare Index}",
  journal = {Geomagnetism and Aeronomy},
     year = 2018,
    month = dec,
   volume = 58,
    pages = {1129-1135},
      doi = {10.1134/S0016793218080078},
   adsurl = {https://ui.adsabs.harvard.edu/abs/2018Ge%26Ae..58.1129F}
}

@ARTICLE{1968SoPh....3..282M,
   author = {{Moreton}, G.~E. and {Severny}, A.~B.},
    title = "{Magnetic Fields and Flares in the Region CMP 20 September 1963}",
  journal = {Solar Physics},
     year = 1968,
    month = feb,
   volume = 3,
    pages = {282-297},
      doi = {10.1007/BF00155163},
   adsurl = {https://ui.adsabs.harvard.edu/abs/1968SoPh....3..282M}
}

@ARTICLE{1990SvA....34..656R,
   author = {{Romanov}, V.~A. and {Tsap}, T.~T.},
    title = "{Magnetic Fields and Flare Activity}",
  journal = {Soviet Astronomy},
     year = 1990,
    month = dec,
   volume = 34,
    pages = {656},
   adsurl = {https://ui.adsabs.harvard.edu/abs/1990SvA....34..656R}
}

@ARTICLE{1990IzKry..81....3A,
   author = {{Abramenko}, V.~I. and {Gopasiuk}, S.~I. and {Ogir'}, M.~B.},
    title = "{Plasma motions and electric currents in an active region}",
  journal = {Izvestiya Ordena Trudovogo Krasnogo Znameni Krymskoj Astrofizicheskoj Observatorii},
     year = 1990,
   volume = 81,
    pages = {3-8},
   adsurl = {https://ui.adsabs.harvard.edu/abs/1990IzKry..81....3A}
}

@ARTICLE{1991RSPTA.336..381C,
   author = {{Canfield}, R.~C. and {de La Beaujardiere}, J.-F. and {Leka}, K.~D.
	},
    title = "{Flare Energy Release: Observational Consequences and Signatures}",
  journal = {Philosophical Transactions of the Royal Society of London Series A},
     year = 1991,
    month = sep,
   volume = 336,
    pages = {381-387},
      doi = {10.1098/rsta.1991.0088},
   adsurl = {https://ui.adsabs.harvard.edu/abs/1991RSPTA.336..381C}
}

@ARTICLE{1997ApJ...482..490L,
   author = {{Li}, J. and {Metcalf}, T.~R. and {Canfield}, R.~C. and {W{\"u}lser}, J.-P. and 
	{Kosugi}, T.},
    title = "{What Is the Spatial Relationship between Hard X-Ray Footpoints and Vertical Electric Currents in Solar Flares?}",
  journal = {The Astrophysical Journal},
     year = 1997,
    month = jun,
   volume = 482,
    pages = {490-497},
      doi = {10.1086/304131},
   adsurl = {https://ui.adsabs.harvard.edu/abs/1997ApJ...482..490L}
}

@ARTICLE{2015A&A...580A.106M,
   author = {{Musset}, S. and {Vilmer}, N. and {Bommier}, V.},
    title = "{Hard X-ray emitting energetic electrons and photospheric electric currents}",
  journal = {Astronomy and Astrophysics},
archivePrefix = "arXiv",
   eprint = {1506.02724},
 primaryClass = "astro-ph.SR",
     year = 2015,
    month = aug,
   volume = 580,
      eid = {A106},
    pages = {A106},
      doi = {10.1051/0004-6361/201424378},
   adsurl = {https://ui.adsabs.harvard.edu/abs/2015A%26A...580A.106M}
}

@ARTICLE{2014ApJ...788L..18S,
   author = {{Sharykin}, I.~N. and {Kosovichev}, A.~G.},
    title = "{Fine Structure of Flare Ribbons and Evolution of Electric Currents}",
  journal = {The Astrophysical Journal Letters},
archivePrefix = "arXiv",
   eprint = {1404.5104},
 primaryClass = "astro-ph.SR",
     year = 2014,
    month = jun,
   volume = 788,
      eid = {L18},
    pages = {L18},
      doi = {10.1088/2041-8205/788/1/L18},
   adsurl = {https://ui.adsabs.harvard.edu/abs/2014ApJ...788L..18S}
}

@ARTICLE{2015ApJ...807..102S,
   author = {{Sharykin}, I.~N. and {Kosovichev}, A.~G. and {Zimovets}, I.~V.
	},
    title = "{Energy Release and Initiation of a Sunquake in a C-class Flare}",
  journal = {The Astrophysical Journal},
archivePrefix = "arXiv",
   eprint = {1405.5912},
 primaryClass = "astro-ph.SR",
     year = 2015,
    month = jul,
   volume = 807,
      eid = {102},
    pages = {102},
      doi = {10.1088/0004-637X/807/1/102},
   adsurl = {https://ui.adsabs.harvard.edu/abs/2015ApJ...807..102S}
}

@ARTICLE{2016ARep...60..939L,
   author = {{Livshits}, M.~A. and {Grigoryeva}, I.~Y. and {Myshyakov}, I.~I. and 
	{Rudenko}, G.~V.},
    title = "{Evidence for a relationship between emerging magnetic fields, electric currents, and solar flares observed on May 10, 2012}",
  journal = {Astronomy Reports},
archivePrefix = "arXiv",
   eprint = {1604.07073},
 primaryClass = "astro-ph.SR",
     year = 2016,
    month = oct,
   volume = 60,
    pages = {939-948},
      doi = {10.1134/S1063772916090031},
   adsurl = {https://ui.adsabs.harvard.edu/abs/2016ARep...60..939L}
}

@ARTICLE{2017AGUFMSH41A2751Z,
   author = {{Zimovets}, I.~V. and {Sharykin}, I.~N. and {Wang}, R. and {Liu}, Y.~D. and 
	{Kosovichev}, A.~G.},
    title = "{Relationship between Hard X-Ray Footpoint Sources and Photospheric Electric Currents in Solar Flares: a Statistical Study}",
  journal = {AGU Fall Meeting Abstracts},
     year = 2017,
    month = dec,
   adsurl = {https://ui.adsabs.harvard.edu/abs/2017AGUFMSH41A2751Z}
}

@ARTICLE{2016PASJ...68..101K,
   author = {{Kang}, J. and {Magara}, T. and {Inoue}, S. and {Kubo}, Y. and 
	{Nishizuka}, N.},
    title = "{Distribution characteristics of coronal electric current density as an indicator for the occurrence of a solar flare}",
  journal = {Publications of the Astronomical Society of Japan},
     year = 2016,
    month = dec,
   volume = 68,
      eid = {101},
    pages = {101},
      doi = {10.1093/pasj/psw092},
   adsurl = {https://ui.adsabs.harvard.edu/abs/2016PASJ...68..101K}
}

@ARTICLE{2016A&A...596A..56M,
   author = {{Moraitis}, K. and {Toutountzi}, A. and {Isliker}, H. and {Georgoulis}, M. and 
	{Vlahos}, L. and {Chintzoglou}, G.},
    title = "{An observationally-driven kinetic approach to coronal heating}",
  journal = {Astronomy and Astrophysics},
archivePrefix = "arXiv",
   eprint = {1603.07129},
 primaryClass = "astro-ph.SR",
     year = 2016,
    month = nov,
   volume = 596,
      eid = {A56},
    pages = {A56},
      doi = {10.1051/0004-6361/201527890},
   adsurl = {https://ui.adsabs.harvard.edu/abs/2016A%26A...596A..56M}
}

@ARTICLE{1919ApJ....49..153H,
   author = {{Hale}, G.~E. and {Ellerman}, F. and {Nicholson}, S.~B. and 
	{Joy}, A.~H.},
    title = "{The Magnetic Polarity of Sun-Spots}",
  journal = {The Astrophysical Journal},
     year = 1919,
    month = apr,
   volume = 49,
    pages = {153},
      doi = {10.1086/142452},
   adsurl = {https://ui.adsabs.harvard.edu/abs/1919ApJ....49..153H}
}

@ARTICLE{2012SoPh..275..207S,
   author = {{Scherrer}, P.~H. and {Schou}, J. and {Bush}, R.~I. and {Kosovichev}, A.~G. and 
	{Bogart}, R.~S. and {Hoeksema}, J.~T. and {Liu}, Y. and {Duvall}, T.~L. and 
	{Zhao}, J. and {Title}, A.~M. and {Schrijver}, C.~J. and {Tarbell}, T.~D. and 
	{Tomczyk}, S.},
    title = "{The Helioseismic and Magnetic Imager (HMI) Investigation for the Solar Dynamics Observatory (SDO)}",
  journal = {Solar Physics},
     year = 2012,
    month = jan,
   volume = 275,
    pages = {207-227},
      doi = {10.1007/s11207-011-9834-2},
   adsurl = {https://ui.adsabs.harvard.edu/abs/2012SoPh..275..207S}
}

@ARTICLE{2002SoPh..210....3L,
   author = {{Lin}, R.~P. and {Dennis}, B.~R. and {Hurford}, G.~J. and {Smith}, D.~M. and 
	{Zehnder}, A. and {Harvey}, P.~R. and {Curtis}, D.~W. and {Pankow}, D. and 
	{Turin}, P. and {Bester}, M. and {Csillaghy}, A. and {Lewis}, M. and 
	{Madden}, N. and {van Beek}, H.~F. and {Appleby}, M. and {Raudorf}, T. and 
	{McTiernan}, J. and {Ramaty}, R. and {Schmahl}, E. and {Schwartz}, R. and 
	{Krucker}, S. and {Abiad}, R. and {Quinn}, T. and {Berg}, P. and 
	{Hashii}, M. and {Sterling}, R. and {Jackson}, R. and {Pratt}, R. and 
	{Campbell}, R.~D. and {Malone}, D. and {Landis}, D. and {Barrington-Leigh}, C.~P. and 
	{Slassi-Sennou}, S. and {Cork}, C. and {Clark}, D. and {Amato}, D. and 
	{Orwig}, L. and {Boyle}, R. and {Banks}, I.~S. and {Shirey}, K. and 
	{Tolbert}, A.~K. and {Zarro}, D. and {Snow}, F. and {Thomsen}, K. and 
	{Henneck}, R. and {McHedlishvili}, A. and {Ming}, P. and {Fivian}, M. and 
	{Jordan}, J. and {Wanner}, R. and {Crubb}, J. and {Preble}, J. and 
	{Matranga}, M. and {Benz}, A. and {Hudson}, H. and {Canfield}, R.~C. and 
	{Holman}, G.~D. and {Crannell}, C. and {Kosugi}, T. and {Emslie}, A.~G. and 
	{Vilmer}, N. and {Brown}, J.~C. and {Johns-Krull}, C. and {Aschwanden}, M. and 
	{Metcalf}, T. and {Conway}, A.},
    title = "{The Reuven Ramaty High-Energy Solar Spectroscopic Imager (RHESSI)}",
  journal = {Solar Physics},
     year = 2002,
    month = nov,
   volume = 210,
    pages = {3-32},
      doi = {10.1023/A:1022428818870},
   adsurl = {https://ui.adsabs.harvard.edu/abs/2002SoPh..210....3L}
}

@ARTICLE{2014SoPh..289.3549B,
   author = {{Bobra}, M.~G. and {Sun}, X. and {Hoeksema}, J.~T. and {Turmon}, M. and 
	{Liu}, Y. and {Hayashi}, K. and {Barnes}, G. and {Leka}, K.~D.
	},
    title = "{The Helioseismic and Magnetic Imager (HMI) Vector Magnetic Field Pipeline: SHARPs - Space-Weather HMI Active Region Patches}",
  journal = {Solar Physics},
archivePrefix = "arXiv",
   eprint = {1404.1879},
 primaryClass = "astro-ph.SR",
     year = 2014,
    month = sep,
   volume = 289,
    pages = {3549-3578},
      doi = {10.1007/s11207-014-0529-3},
   adsurl = {https://ui.adsabs.harvard.edu/abs/2014SoPh..289.3549B}
}

@ARTICLE{2014SoPh..289.3483H,
   author = {{Hoeksema}, J.~T. and {Liu}, Y. and {Hayashi}, K. and {Sun}, X. and 
	{Schou}, J. and {Couvidat}, S. and {Norton}, A. and {Bobra}, M. and 
	{Centeno}, R. and {Leka}, K.~D. and {Barnes}, G. and {Turmon}, M.
	},
    title = "{The Helioseismic and Magnetic Imager (HMI) Vector Magnetic Field Pipeline: Overview and Performance}",
  journal = {Solar Physics},
archivePrefix = "arXiv",
   eprint = {1404.1881},
 primaryClass = "astro-ph.SR",
     year = 2014,
    month = sep,
   volume = 289,
    pages = {3483-3530},
      doi = {10.1007/s11207-014-0516-8},
   adsurl = {https://ui.adsabs.harvard.edu/abs/2014SoPh..289.3483H}
}

@ARTICLE{2002A&A...395.1077C,
   author = {{Calabretta}, M.~R. and {Greisen}, E.~W.},
    title = "{Representations of celestial coordinates in FITS}",
  journal = {Astronomy and Astrophysics},
   eprint = {astro-ph/0207413},
     year = 2002,
    month = dec,
   volume = 395,
    pages = {1077-1122},
      doi = {10.1051/0004-6361:20021327},
   adsurl = {https://ui.adsabs.harvard.edu/abs/2002A%26A...395.1077C}
}

@ARTICLE{2006A&A...449..791T,
   author = {{Thompson}, W.~T.},
    title = "{Coordinate systems for solar image data}",
  journal = {Astronomy and Astrophysics},
     year = 2006,
    month = apr,
   volume = 449,
    pages = {791-803},
      doi = {10.1051/0004-6361:20054262},
   adsurl = {https://ui.adsabs.harvard.edu/abs/2006A%26A...449..791T}
}

@ARTICLE{2019arXiv190610788S,
   author = {{Sadykov}, V.~M and {Kosovichev}, A.~G and {Kitiashvili}, I.~N and 
	{Kerr}, G.~S},
    title = "{Response of SDO/HMI observables to heating of the solar atmosphere by precipitating high-energy electrons}",
  journal = {arXiv e-prints},
archivePrefix = "arXiv",
   eprint = {1906.10788},
 primaryClass = "astro-ph.SR",
     year = 2019,
    month = jun,
   adsurl = {https://ui.adsabs.harvard.edu/abs/2019arXiv190610788S}
}

@ARTICLE{2019LRSP...16....3T,
   author = {{Toriumi}, S. and {Wang}, H.},
    title = "{Flare-productive active regions}",
  journal = {Living Reviews in Solar Physics},
archivePrefix = "arXiv",
   eprint = {1904.12027},
 primaryClass = "astro-ph.SR",
     year = 2019,
    month = may,
   volume = 16,
      eid = {3},
    pages = {3},
      doi = {10.1007/s41116-019-0019-7},
   adsurl = {https://ui.adsabs.harvard.edu/abs/2019LRSP...16....3T}
}

@ARTICLE{2005ApJ...619.1160A,
   author = {{Abramenko}, V.~I. and {Longcope}, D.~W.},
    title = "{Distribution of the Magnetic Flux in Elements of the Magnetic Field in Active Regions}",
  journal = {The Astrophysical Journal},
     year = 2005,
    month = feb,
   volume = 619,
    pages = {1160-1166},
      doi = {10.1086/426710},
   adsurl = {https://ui.adsabs.harvard.edu/abs/2005ApJ...619.1160A}
}

@ARTICLE{2005ApJ...629.1141A,
   author = {{Abramenko}, V.~I.},
    title = "{Relationship between Magnetic Power Spectrum and Flare Productivity in Solar Active Regions}",
  journal = {The Astrophysical Journal},
     year = 2005,
    month = aug,
   volume = 629,
    pages = {1141-1149},
      doi = {10.1086/431732},
   adsurl = {https://ui.adsabs.harvard.edu/abs/2005ApJ...629.1141A}
}

@ARTICLE{2010ApJ...720..717A,
   author = {{Abramenko}, V. and {Yurchyshyn}, V.},
    title = "{Magnetic Energy Spectra in Solar Active Regions}",
  journal = {The Astrophysical Journal},
archivePrefix = "arXiv",
   eprint = {1007.3702},
 primaryClass = "astro-ph.SR",
     year = 2010,
    month = sep,
   volume = 720,
    pages = {717-722},
      doi = {10.1088/0004-637X/720/1/717},
   adsurl = {https://ui.adsabs.harvard.edu/abs/2010ApJ...720..717A}
}

@ARTICLE{2018MNRAS.480.3780K,
   author = {{Kutsenko}, A.~S. and {Abramenko}, V.~I. and {Kuzanyan}, K.~M. and 
	{Xu}, H. and {Zhang}, H.},
    title = "{Intermittency spectra of current helicity in solar active regions}",
  journal = {Monthly Notices of the Royal Astronomical Society},
archivePrefix = "arXiv",
   eprint = {1802.02323},
 primaryClass = "astro-ph.SR",
     year = 2018,
    month = nov,
   volume = 480,
    pages = {3780-3787},
      doi = {10.1093/mnras/sty2109},
   adsurl = {https://ui.adsabs.harvard.edu/abs/2018MNRAS.480.3780K}
}
\end{document}